\documentstyle[12pt, a4]{article}
\begin{document}
\author{M Ko\c{c}ak and B G\"{o}n\"{u}l
\and Department of Engineering Physics, Faculty of Engineering,
\and University of Gaziantep, 27310 Gaziantep -Turkey}
\title{A search on Dirac equation }
\date{}
\maketitle

The solutions, in terms of orthogonal polynomials, of  Dirac
equation with analytically solvable potentials are investigated
within a novel formalism by transforming the relativistic equation
into a Schr\"{o}dinger like one. Earlier results are discussed in
a unified framework and certain solutions of a large class of
potentials are given.
\newline


\newline

\section{Introduction}
A new algebraic technique for solving Schr\"{o}dinger  and
Klein-Gordon equations \cite{gonul1}, and the related other works
therein, has been introduced recently and used to search many
interesting problems in different disciplines of physics. These
works have clarified the power of the suggested model when
compared the results obtained with those provided by other
analytical methods in the literature. Nevertheless, this formalism
involves a deficiency in its present form which requires, to
express the excited state wave functions, the application of
linear operators on the ground state wave function appeared
automatically in the mathematical framework. This is indeed a
cumbersome procedure though it provides explicit expressions for
the state functions having non-zero angular momenta.

To remove this drawback inherent in the formalism used in our
previous works \cite{gonul1}, we suggest here an alternative
scheme, unifying the spirit of the two theoretical models
\cite{gonul1, nikiforov}, to work out
relativistic/non-relativistic quantum mechanical problems
analytically in a unified framework. This is the main motivation
behind the work presented in this article which in particular
focuses on the solution of Dirac equation since recently
considerable attention has been paid to exactly solvable Dirac
equations.

The arrangement of this article is as follows. In the next
section, a brief introduction of the usual Dirac formalism and its
treatment within the frame of new scheme are presented. Third
section involves application results. Finally, the results are
summarized in the concluding section.

\section{Theoretical Consideration}

To proceed, let us first remind the mathematical frame of Dirac
formalism which is discussed briefly in the following section.
Section 2.2 then illustrates the formalism of the new model.

\subsection{Background on Dirac equation}

Dirac equation for scalar and vector potentials is given by
\cite{greiner} ($\hbar=c=1$)
\begin{equation}
H\Psi=\{\hat{\alpha}.\hat{p}+\beta(m+V_{S})+V_{V}\}\Psi
\end{equation}
where $\hat{p}$ is the momentum operator, $m$ is the rest mass of
the particle, $V_{S}$ and $V_{V}$ are scalar and vector potentials
respectively and $\hat{\alpha}$, $\beta$ are Pauli matrices.

To separate angular part of Eq. (1) from the radial part one
follows
\begin{equation}
\Psi^{\ell}_{j m}=\left[
\begin{array}{c}
\frac{iG_{\ell j}}{r}\varphi^{\ell}_{j m}\\
\frac{F_{\ell
j}}{r}\frac{\hat{\sigma}.\hat{r}}{r}\varphi^{\ell}_{j m}
\end{array}
\right]
\end{equation}
where $\hat{\sigma}$ represents the Pauli spin matrices while
$G_{\ell j}$ and $F_{\ell j}$ are the radial components
\newline
~~~~~$G_{\ell j}=\left\{
\begin{array}{c}
G^{+}_{j}~~~j=\ell+1/2\\
G^{-}_{j}~~~j=\ell-1/2
\end{array}\right\}~~~~~,
~~~~~F_{\ell j}=\left\{
\begin{array}{c}
F^{+}_{j}~~~j=\ell+1/2\\
F^{-}_{j}~~~~j=\ell-1/2
\end{array}
\right\}$
\newline
and $ \varphi_{j m}^{\ell}$ is the angular part of the wave
function

~~~~~~~~~~~~~~~~~~~~~~~~~~$\varphi_{j m}^{\ell}=\left\{
\begin{array}{c}
\varphi^{+}_{j m}~~~j=\ell+1/2\\
\varphi^{-}_{j m}~~~j=\ell-1/2
\end{array}
\right\}$~~~.
\newline
Then, using the standard relations
\begin{equation}
\hat{\sigma}.\hat{p}\frac{g(r)}{r}\varphi_{j
m}^{\ell}=-\frac{i}{r}(\frac{dg}{dr}+\frac{kg}{r})\frac{\hat{\sigma}.\hat{r}}{r^{2}}\varphi_{j
m}^{\ell}
\end{equation}
and
\begin{equation}
\hat{\sigma}.\hat{p}
\frac{\hat{\sigma}.\hat{r}}{r^{2}}\frac{g(r)}{r}\varphi_{j
m}^{\ell}=-\frac{i}{r}(\frac{dg}{dr}-\frac{kg}{r})\varphi_{j
m}^{\ell}
\end{equation}
where
\newline
~~~~~~$k=\left\{
\begin{array}{c}
-(\ell+1)=-(j+1/2)~~~~j=\ell+1/2\\
+\ell    =+(j+1/2)~~~~~~~~~~j=\ell-1/2
\end{array}
\right\}$ and $g(r)$ is an arbitrary function.
\newline
Hence, the radial equations appear as
\begin{equation}
-\frac{dF(r)}{dr}+\frac{k}{r}F(r)=(\varepsilon-m-V_{S}-V_{V})G(r)
\end{equation}
\begin{equation}
\frac{dG(r)}{dr}+\frac{k}{r}G(r)=(\varepsilon+m+V_{S}-V_{V})F(r),
\end{equation}
where $\varepsilon$ is the total relativistic energy of the
system. From (5) and (6), omitting the derivatives of $V_{S}$ and
$V_{V}$, together with the elimination of one radial component
($F(r)$), one obtains a Schr\"{o}dinger like equation for the
other component of the relativistic wave function
\begin{equation}
\left\{-\frac{d^{2}}{dr^{2}}+\frac{k(k+1)}{r^{2}}+(V^{2}_{S}-V^{2}_{V})+
(2mV_{S}+2\varepsilon V_{V})\right\}G=(\varepsilon^{2}-m^{2})G.
\end{equation}

\subsection{Formalism}
Exact solutions of systems in physics has a great importance. To
provide such solutions recently a new method has been  carried out
successfully in our earlier works \cite{gonul1}, which
unfortunately has a considerable algebraic difficulty in the
calculation of excited state functions. To overcome this tedious
procedure in the calculations, we propose here to use a new scheme
involving orthogonal polynomials in order to express all bound
state wave functions in an explicit form.

For this purpose, we start from Eq.(7) which can be defined as

\begin{equation}
\frac{G''(r)}{G(r)}=V(r)-E~,
\end{equation}
where $V(r)=\frac{k(k+1)}{r^{2}}+(V^{2}_{S}-V^{2}_{V})+
(2mV_{S}+2\varepsilon V_{V})$ and $E=\varepsilon^{2}-m^{2}$. As is
well known, the solution of (8) generally takes the form
\begin{equation}
G(r)=f(r)F[s(r)].
\end{equation}
The substitution of (9) into (8) yields the second-order
differential equation
\begin{equation}
\left(\frac{f''}{f}+\frac{F''s'^{2}}{F}+\frac{s''F'}{F}+\frac{2F's'f'}{Ff}\right)=V-E~,
\end{equation}
and rearranging (10) for a more useful form, one gets
\begin{equation}
F''+\left(\frac{s''}{s'^{2}}+2\frac{f'}{s'f}\right)F'+\left(\frac{f''}{s'^{2}f}+\frac{E-V}{s'^{2}}\right)F=0~.
\end{equation}

Eq. (11) is in the form of the most familiar second-order
differential equations to the hypergeometric type
\cite{nikiforov},
\begin{equation}
F''(s)+\frac{\tau(s)}{\sigma(s)}F'(s)+
\frac{\tilde{\sigma}(s)}{\sigma^{2}(s)}F(s)=0~,
\end{equation}
where $\sigma$ and $\tilde{\sigma}$ are at most second degree
polynomials, and $\tau$ is a first degree polynomial. The form of
$\frac{\tau(s)}{\sigma(s)}$ and
$\frac{\tilde{\sigma}(s)}{\sigma^{2}(s)}$ is well defined for any
special function $F(s)$ \cite{nikiforov}. From (12), it follows
that
\begin{equation}
\frac{s''}{s'^{2}}+2\frac{f'}{s'f}=\frac{\tau(s)}{\sigma(s)}~,~~~~~~\frac{f''}{fs'^{2}}+\frac{E-V}{s'^{2}}=\frac{\tilde{\sigma}}{\sigma^{2}}~.
\end{equation}
From the previous works in \cite{gonul1}, the energy and potential
terms in (13) can be decomposed in two pieces, which provides a
clear understanding for the individual contributions of the $F$
and $f$ terms to the whole of the solutions, such that
$E-V=(E_{f}+E_{F})-(V_{f}+V_{F})$. Therefore, the second equality
in (13) is transformed to a couple of equation
\begin{equation}
\frac{f''}{f}=V_{f}-E_{f}~,~~~~~~~~-\frac{\tilde{\sigma}
}{\sigma^{2}}s'^{2}=V_{F}-E_{F}~,
\end{equation}
where $f$ can be expressed in an explicit form due to the first
part in (13)
\begin{equation}
f(r)=(s')^{-1/2}\exp{\left[\frac{1}{2}\int^{s(r)}\frac{\tau(s)}{\sigma(s)}ds\right]}~.
\end{equation}
Since the corresponding $\sigma,~ \tilde{\sigma}$ and $\tau$ terms
are well known for a given polynomial $(F)$, the transformation
function $(s)$ in (14), and afterwards $f$ in (15), are easily
defined. So, from (9), the corresponding total wave function is
readily obtained for the whole spectrum.

The potential and total energy terms for Dirac equation in this
case
\begin{equation}
\frac{f''}{f}=V_{f}-E_{f} ,~~~~~~~~V_{f}=2mV_{S}+2\varepsilon
V_{V}+\frac{k(k+1)}{r^{2}}~,
\end{equation}
\begin{equation}
-\frac{\tilde{\sigma} }{\sigma^{2}}s'^{2}=V_{F}-E_{F}
,~~~~~~~~V_{F}=V_{S}^{2}-V_{V}^{2}~,
\end{equation}
and
\begin{equation}
E_{f}+E_{F}=\varepsilon^{2}-m^{2}~.
\end{equation}

To understand how efficiently this method works, some physically
possible potentials are solved in the following section to obtain
their eigenvalues and eigenfunctions, within the frame of the
present formalism.

\section{Application}

As illustrative examples, here we deal with the two well known
problems of the literature: Dirac oscillator and Dirac-Coulomb
problem. The other solutions are shown in Table 1.

In order to get exact solutions for the present consideration, one
needs to start with choosing a physically plausible equal
magnitudes for the vector and scalar potentials. Otherwise, the
system considered becomes quasi-exactly solvable which is out of
the scope of the present article.

Setting $V_{V}=V_{S}=ar^{2}$ for the relativistic treatment of
oscillator problem right hand sides of equations (16) and (17)
gives
\begin{equation}
V_{f}=2a(m+\varepsilon)r^{2}+\frac{k(k+1)}{r^{2}}~,~~~~~~~~~V_{F}=0~,
\end{equation}

Concentrating on the generalized Laguerre polynomials
$L_{n}^{\alpha}(s)$ related to confluent hypergeometric functions,
one sees that \cite{nikiforov}
\begin{equation}
\sigma=s~,~~~~\tau=\alpha+1-s~,~~~~\tilde{\sigma}=n \sigma~.
\end{equation}
Keeping in mind that the right-hand sides of (14) provide a three
dimensional harmonic oscillator potential, one obviously realizes
that $\frac{s'^{2}}{s}=2w$ and consequently $s=\frac{1}{2}\omega
r^{2}$. Then, substituting (20) into (15) it is not hard to see
that
\begin{equation}
f=Cr^{\alpha+1/2}e^{\frac{-\omega r^{2}}{4}}~,
\end{equation}
where $C=\frac{1}{\sqrt{2}}(\frac{\omega}{2})^{\frac{\alpha}{2}}$.
This makes possible to predict $V_{f}$ and $E_{f}$ as
\begin{eqnarray}
\frac{f''}{f}=V_{f}-E_{f}~,~~~~~~ E_{f}=(\alpha+1)\omega
\nonumber~,
\end{eqnarray}
\begin{eqnarray}
V_{f}=\frac{1}{4}\omega^{2}r^{2}+\frac{(\alpha-1/2)(\alpha+1/2)}{r^{2}}~,
\end{eqnarray}
where $\alpha=-(k+1/2)=\ell+\frac{1}{2}$ for the case
$j=\ell+\frac{1}{2}$. To find also $V_{F}$ and $E_{F}$, one should
consider Eq. (17). After some simple algebra we find
\begin{equation}
V_{F}=0~,~~~~E_{F}=2n \omega~.
\end{equation}
Thus, in the non-relativistic limit, the full energy spectrum and
wave functions for the system of interest are gives as
\begin{eqnarray}
E=E_{f}+E_{F}=(\alpha+1+2n)\omega=(2n+\ell+3/2)\omega \nonumber ~,
\end{eqnarray}
\begin{eqnarray}
\Psi=fF=Cs^{\frac{(\ell+1)}{2}}e^{-\frac{s}{2}}L^{(\ell+\frac{1}{2})}_{n}(s)~.
\end{eqnarray}
Finally, the relativistic energy of Dirac oscillator reads
\begin{equation}
\varepsilon^{2}=m^{2}+(2n+\ell+3/2)\omega.
\end{equation}
The results obtained are in agreement with the work of Levai
\cite{nikiforov} which considers only the non-relativistic case,
and also, for proper parameters ($\ell+1=\kappa$,
$w=4\alpha^{2}\zeta$), these results agree well with the study of
Alhaidari \cite{alhaidari1}. Moreover, the findings justify the
excellent discussion in \cite{keng-su} on the confinement
properties for Dirac equation with scalar and vector like
potentials.

Obviously, from the similarity between Eqs. (19) and (22) it is
clear that $a=\frac{w^{2}}{8(m+\varepsilon)}$.

It is importantly noted that the choose of equal magnitudes for
vector and scalar potentials leads to the non-relativistic limit
of Dirac equation, removing the relativistic corrections. For a
comprehensive understanding of this interesting point, the reader
is referred to the individual works of G\"{o}n\"{u}l and Ko\c{c}ak
in \cite{gonul1} regarding the treatment of Klein-Gordon equation.
We additionally remark that the present algebraic treatment has
been performed only for spin-up case. Clearly, following similar
procedure, one can easily repeat the same calculations for the
other spinor wave function where now $k=+\ell$.

The relativistic hydrogen atom is also an exactly solvable system
within the frame of Dirac equation where the piece of potentials
now should be defined as
\begin{equation}
\frac{f''}{f}=V_{f}-E_{f} ,~~~~~~~~V_{f}=V_{S}^{2}-V_{V}^{2}~,
\end{equation}
\begin{equation} -\frac{\tilde{\sigma}}{\sigma^{2}}s'^{2}=V_{F}-E_{F}
,~~~~~~~~V_{F}=2mV_{S}+2\varepsilon V_{V}+\frac{k(k+1)}{r^{2}}~.
\end{equation}
For again the equal vector and scalar potentials;
$V_{V}=V_{S}=-\frac{b}{r}$ one gets
\begin{equation}
V_{f}=0~,~~~~~~~~V_{F}=-\frac{2(m+\varepsilon)b}{r}+\frac{k(k+1)}{r^{2}}~.
\end{equation}
In order to apply the present orthogonal polynomial technique, we
choose the most suitable generalized Laguerre polynomial
[$F=e^{-s/2}s^{\frac{\alpha+1}{2}}L^{\alpha}_{n}(s)$] where
\begin{equation}
\sigma=1~,~~\tau=0~,~~\tilde{\sigma}=\frac{2n+\alpha+1}{2s}+\frac{1-\alpha^{2}}{4s^{2}}-\frac{1}{4},
\end{equation}
leading to $s=ar$, and to be in convenience with the literature,
we set $a=\frac{e^{2}}{n+\ell+1}$. Then, Eq. (15) reads
\begin{equation}
f=\frac{(n+\ell+1)^{\frac{1}{2}}}{e}.
\end{equation}
This justifies that $\frac{f''}{f}=V_{f}-E_{f}=0$, while
\begin{equation}
V_{F}=-\frac{e^{2}}{r}+\frac{\ell(\ell+1)}{r^{2}}~,~~~~~E_{F}=-\frac{e^{4}}{4(n+\ell+1)^{2}}~,
\end{equation}
where $k=-(\ell+1)$ and $\alpha=2\ell+1$. Comparing Eqs. (28) and
(31) one sees that $b=\frac{e^{2}}{2(m+\varepsilon)}$. Thus, for
this system, the full energy spectrum and wave functions are
\begin{eqnarray}
E=E_{f}+E_{F}=E_{F}=-\frac{e^{4}}{4(n+\ell+1)^{2}}~,
\end{eqnarray}
\begin{eqnarray}
\Psi=fF=Ce^{-s/2}s^{\ell+1}L^{2\ell+1}_{n}(s),
\end{eqnarray}
where $C=\frac{(n+\ell+1)^{\frac{1}{2}}}{e}$. Finally, the
relativistic energy for Dirac-Coulomb problem is
\begin{eqnarray}
\varepsilon^{2}=m^{2}-\frac{e^{4}}{4(n+\ell+1)^{2}}~~~.
\end{eqnarray}
The results are in agreement with the work of Levai
\cite{nikiforov} and, with a suitable parameters such that
$1-\left[\frac{e^{2}}{2(n+\ell+1)}\right]^{2}=\left[1+\left(\frac{\alpha
Z}{\gamma+n+1}\right)^{2}\right]^{-1}$, overlap with those
obtained by Alhaidari in \cite{alhaidari1}.

The formalism used here also provides explicit expressions for the
relativistic spectra of three more potentials, which are
illustrated in Table 1. We stress that, unlike the work in
\cite{alhaidari1}, the strategy followed in this article for
transforming Dirac equation into a Schr\"{o}dinger like one does
not enlarge the class of exactly solvable potentials as much as it
might appear at first sight. Hence, the present brief work
supports the individual criticism of the Castro and
Vaidya-Rodrigues \cite{castro}.

It is additionally stressed that one of the potentials listed in
Table 1 do not require a restriction as in the examples discussed
above in defining equal magnitudes for vector and scalar
potentials leading to the exact solvability.

\section{Concluding Remarks}

We have presented an idea of connecting the methods used in the
analysis of exactly solvable potentials in non-relativistic
quantum mechanics with the solution procedure of Dirac equation.
The suggested formalism systematically recovers known results in a
natural unified way and allows one to extend certain results known
in particular cases. A straightforward generalization would be the
application of the scheme to the other relativistic equations for
integer spin cases. Beyond its intrinsic importance as a new
solution for a fundamental equation in physics, we also expect
that the present simple method would find a widespread application
in the study of different quantum mechanical and nuclear
scattering systems. Along this line the works are in progress.

\newpage
\begin{table}
\begin{tabular}{|l|l|l|l|}
\hline
 & $Oscillator$ & $Coulomb$ & $Morse$  \\
\hline
$V_{S}$ & $ar^{2}$ & $-b/r$ & $-Ae^{-ar}+(\sqrt{B^{2}+m^{2}}-m)$  \\
\hline
$V_{V}$ & $ar^{2}$ & $-b/r$ & $-Ce^{-ar}$  \\
\hline
$s$& $\frac{1}{2}\omega r^{2}$ & $(e^{2}/(n+\ell+1))r$ & $(2D/a)e^{-ax}~;~~D^{2}=A^{2}-C^{2}$ \\
\hline
$\sigma$ & $s$ & $1$ & $s$  \\
\hline
$\tau$ & $\alpha+1-s$ & $0$ & $\alpha+1-s$ \\
\hline $\tilde{\sigma}$ & $ns$ &
$\frac{2n+\alpha+1}{2s}+\frac{1-\alpha^{2}}{4s^{2}}-\frac{1}{4}$ &
$ns$  \\
\hline
$\varepsilon$ & $\left(m^{2}+(2n+\ell+3/2)\omega\right)^{\frac{1}{2}}$  & $\left(m^{2}-\frac{e^{4}}{4(n+\ell+1)^{2}}\right)^{\frac{1}{2}}$ & $\left(m^{2}+B^{2}-\frac{a^{2}\alpha^{2}}{4}\right)^{\frac{1}{2}}$ \\
\hline $\Psi$ &
$s^{\frac{(2\alpha+1)}{4}}e^{-\frac{s}{2}}L^{(\alpha)}_{n}(s)$ &
$s^{\frac{\alpha+1}{2}}e^{-s/2}L^{\alpha}_{n}(s)$ &
$s^{\alpha/2}e^{-\frac{s}{2}}L_{n}^{\alpha}(s)$ \\
\hline
$\alpha$ & $\ell+1/2$ & $2\ell+1$ & $\frac{2(A\sqrt{B^{2}+m^{2}}+\varepsilon C)}{a\sqrt{A^{2}-C^{2}}}-1-2n$ \\
\hline
\end{tabular}

\begin{tabular}{|l|l|l|}
\hline
 & $Rosen-Morse$ & $Eckart$  \\
\hline
$V_{S}$ & $(A\tanh(ar)+B)^{2}$ & $(-A\coth(ar)+B)^{2}$ \\
\hline
$V_{V}$ & $(A\tanh(ar)+B)^{2}$ & $(-A\coth(ar)+B)^{2}$ \\
\hline
$s$ & $\tanh{ar}$ & $\coth{ar}$ \\
\hline $\sigma$  & $1$ &
$1$ \\
\hline
$\tau$ &  $0$ & $0$ \\
\hline $\tilde{\sigma}$ &
$\frac{1-\alpha^{2}}{4(1-s^{2})}+\frac{1-\beta^{2}}{4(1+s^{2})}+\frac{c_{n}}{(1-s^{2})}$
& $\frac{1-\alpha^{2}}{4(1-s^{2})}+\frac{1-\beta^{2}}{4(1+s^{2})}
+\frac{c_{n}}{(1-s^{2})}$ \\
\hline
$\varepsilon$ & $\left(m^{2}+\eta-\frac{a^{2}(\alpha^{2}+\beta^{2})}{2}\right)^{\frac{1}{2}}$ & $\left(m^{2}+\zeta-\frac{a^{2}(\alpha^{2}+\beta^{2})}{2}\right)^{\frac{1}{2}}$ \\
\hline
$\Psi$ & $(1-s)^{\alpha/2}(1+s)^{\beta/2}P_{n}^{\alpha,\beta}(s)$ & $(s-1)^{\alpha/2}(1+s)^{\beta/2}P_{n}^{\alpha,\beta}(s)$ \\
\hline $\alpha$ & $\gamma-n+\frac{\lambda}{\gamma-n}$ &
 $-\gamma-n+\frac{\lambda}{\gamma+n}$ \\
\hline
$\beta$ & $\gamma-n-\frac{\lambda}{\gamma-n}$& $-\gamma-n-\frac{\lambda}{\gamma+n}$\\
\hline $c_{n}$ &
$n(n+\alpha+\beta+1)+\frac{1}{2}(\alpha+1)(\beta+1)$ & $n(n+\alpha+\beta+1)+\frac{1}{2}(\alpha+1)(\beta+1)$ \\
\hline
\end{tabular}
\label{table1.1} \caption{Relativistic energy and unnormalized
eigenfunctions of the five potentials deduced within the present
Dirac formalism discussed in section 3. In the treatment of
Rosen-Morse and Eckart potentials, the notation carried out in
\cite{dabrowska} is used.}
\end{table}


\newpage
\begin{thebibliography}{99}

\bibitem{gonul1} B G\"{o}n\"{u}l, K K\"{o}ksal  \textit{Phys. Scr.} \textbf{73} (2006) 629;
B G\"{o}n\"{u}l, K K\"{o}ksal and E Bakir \textit{Phys. Scr.}
\textbf{73} (2006) 279; B G\"{o}n\"{u}l \textit{Chinese
Phys.Lett.} \textbf{23} (2006) 2640; M Ko\c{c}ak \textit{Chinese
Phys.Lett.} \textbf{24} (2007) 315.

\bibitem{nikiforov} A F Nikiforov and V B Uvarov \emph{Special Functions of
Mathematical Physics} (Basle: Birkhauser) 1988; G Levai 1989
\textit{J. Phys.} \textbf{A22} (1989) 689.

\bibitem{greiner} W Greiner; \emph{Relativistic Quantum Mechanics}, third ed.,
(Springer, Berlin) 2000; J D Bjorken and S. D. Drell
\emph{Relativistic Quantum Mechanics} (McGraw-Hill Book Company,
New York) 1964.

\bibitem{alhaidari1} A D Alhaidari \textit{Phys. Rev. Lett.} \textbf{87}
(2001) 210405; A D Alhaidari \textit{J. Phys.} \textbf{A34} (2001)
9827.

\bibitem{keng-su} R K Su, Z Q Ma  \textit{J. Phys.} \textbf{A19} (1986)
1739.

\bibitem{castro} A S de Castro \textit{J. Phys.} \textbf{A35}
(2002)6203; A N Vaidya, R L Rodrigues, hep-th/0203067.

\bibitem{dabrowska} J W Dabrowska, A Khare and U Sukhatme \textit{J. Phys.}
\textbf{A21} (1988) L195.
\end{thebibliography}
\end{document}